\documentclass[Times,16pt,aps,pra,onecolumn,showpacs,amsmath,amssymb,floatfix,footinbib,superscriptaddress,times]{revtex4-1}
\usepackage{graphics,graphicx,epsfig,ulem,epstopdf,bm,longtable,url,datetime}
\usepackage[colorlinks=true,linkcolor=blue,urlcolor=blue,citecolor=blue]{hyperref}
\usepackage{amsmath,amssymb,latexsym,float,amsfonts,subfigure,hyperref,color}
\usepackage[ansinew]{inputenc}
\usepackage[usenames,dvipsnames]{pstricks}
\usepackage[mathlines]{lineno}
\def\footnoterule{\kern -1mm \hrule width 5.8cm \kern 2.2mm}%
\usepackage{tikz,xcolor,hyperref}
\definecolor{lime}{HTML}{A6CE39}
\DeclareRobustCommand{\orcidicon}{%
    \begin{tikzpicture}
    \draw[lime, fill=lime] (0,0)
    circle [radius=0.16]
    node[white] {{\fontfamily{qag}\selectfont \tiny ID}};\draw[white, fill=white] (-0.0625,0.095)
    circle [radius=0.007];
    \end{tikzpicture}
    \hspace{-2mm}}
\foreach \x in {A, ..., Z}
{\expandafter\xdef\csname orcid\x\endcsname{\noexpand\href{https://orcid.org/\csname orcidauthor\x\endcsname}{\noexpand\orcidicon}}}

\begin{document}
\title{Algebraic analysis of electromagnetic chirality-induced negative refractive index in a four-level atomic system}

\author{Shuncai Zhao \orcidA{}}%
\email[Corresponding author: ]{zhaosc@kmust.edu.cn.}
\affiliation{Physics department, Kunming University of Science and Technology, Kunming, 650500, PR China}

\author{Qi-Xuan Wu}
\affiliation{College English department,Kunming University of Science and Technology, Kunming,650500,PR China}

\author{Ai-Ling Gong}%
\affiliation{Physics department, Kunming University of Science and Technology, Kunming, 650500, PR China}
\begin{abstract}
This paper presents a algebraic analysis of
electromagnetic chirality-induced negative refractive index in a
four-level atomic medium. According to analyze mathematically its
argument of the complex refractive index for one circular
polarization, it found that the negative refractive index without
simultaneously negative permittivity and permeability can be
obtained when the argument is in the second quadrant of the
cartesian coordinate system, and that the probe field coupling to
two equal transition frequencies in the atomic level doesn't
require. This undoubtedly reduced stringent conditions to negative
refractive index by quantum optics. As an application, our scheme
may possibly give a novel approach to obtain negative refractive
index by electromagnetic chirality-inducing.
\end{abstract}

\maketitle
\section{Introduction}
Negative refraction is an intriguing and
counterintuitive phenomenon, and impressive efforts have recently
been made to investigate negative refractive index
materials\cite{Ref1}-\cite{Ref3}. Materials with negative refractive
index promise many surprising properties: in the original
description\cite{Ref4}, it was stated that materials with negative
refractive index promise many surprising and even counterintuitive
electromagnetical and optical effects, such as the reversals of both
Doppler shift and Cerenkov radiation\cite{Ref3}, amplification of
evanescent waves\cite{Ref5},subwavelength
focusing\cite{Ref5}-\cite{Ref7}and so on. Up to now, there have been
several approaches to the realization of negative refractive index
materials, including artificial composite
metamaterials\cite{Ref1},\cite{Ref9}, photonic crystal
structures\cite{Ref10}, transmission line simulation\cite{Ref11}and
chiral media\cite{Ref12}-\cite{Ref13}as well as photonic resonant
materials(coherent atomic vapour)[13]-[18]. And the early proposals
for negative refraction required media with both negative
permittivity and permeability ($\varepsilon,\mu$$<$0) in the
frequency range of interest\cite{Ref4}. However, the typical
magnetic dipole transition is smaller than electric dipole
transition by a factor of the order of the fine structure
constant($\alpha\approx\frac{1}{137}$). It is difficult to achieve
negative permeability. An elegant scheme for negative refraction is
proposed in the electromagnetic chirality medium without
simultaneously negative permittivity and
permeability\cite{Ref12},\cite{Ref17}-\cite{Ref29}. A medium in
which the electric polarization \textbf{P} is coupled to the
magnetic field \textbf{H} of an electromagnetic wave and the
magnetization \textbf{M} is coupled to the electric field
\textbf{E}\cite{Ref12}:
\begin{equation}
\textbf{P}=\varepsilon_{0}\chi_{e}\textbf{E}+\frac{\xi_{EH}}{c}\textbf{H},\nonumber\\
\textbf{M}=\frac{\xi_{HE}}{c \mu_{0}}\textbf{E}+\chi_{m}\textbf{H}
\end{equation}
Here$\chi_{e}$and$\chi_{m}$, and $\xi_{EH}$ and $\xi_{HE}$ are the
electric and magnetic susceptibilities, and the complex chirality
coefficients, respectively.They lead to additional contributions to
the refractive index for one circular
polarization\cite{Ref12},\cite{Ref18},\cite{Ref19}:
\begin{equation}
n=\sqrt{\varepsilon\mu-\frac{(\xi_{EH}+\xi_{HE})^{2}}{4}}+\frac{i}{2}(\xi_{EH}-\xi_{HE})
\end{equation}
Obviously, if $\xi_{EH}=-\xi_{HE}=i\xi$,the real part of
$\sqrt{\varepsilon\mu}$ is less than $\xi$($\xi>0$), the negative
refraction is obtained without requiring both $\varepsilon<0$ and
$\mu < 0$. In this paper, we analyze the argument of the complex
number [$\varepsilon\mu-\frac{(\xi_{EH}+\xi_{HE})^{2}}{4}$], and get
the conclusion that when its argument is in the second quadrant of
the cartesian coordinate system, the argument of its square root
[$\sqrt{\varepsilon\mu-\frac{(\xi_{EH}+\xi_{HE})^{2}}{4}}$]has the
possibility in the third quadrant accompanying with negative real
part. Then the negative refractive index can be realized under the
appropriate conditions without the stringent simultaneously negative
permittivity and permeability.

The paper is organized as follows. In Section 2, we present our
model and its expressions for the chirality coefficients and
refractive index. In Section 3, we present numerical results and
their discussion. This is followed by concluding remarks in Section
4.

\section{Model and chirality coefficients}

\begin{figure}
\centering
\resizebox{0.20\textwidth}{!}{%
  \includegraphics{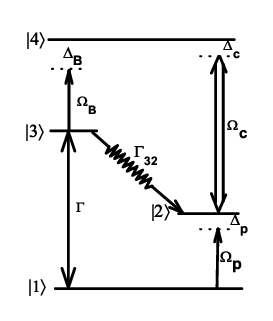}
  }
  \caption{Schematic diagram of a four-level atomic system
interacting with a coherent $\Omega_{c}$ , an incoherent pumps
$\Gamma$ and a probe field which electric and magnetic components
are coupled to the level pairs$|2\rangle-|1\rangle and
|3\rangle-|4\rangle$,respectively. }
\label{fig:1}
\end{figure}

In the following we mathematically analysis electromagnetically
induced chiral negative refraction. The four-level configuration of
atoms for consideration is shown in Figure 1. The parity properties
of the atomic states are as following: levels $|1\rangle$,
$|3\rangle$, and $|4\rangle$ have same parity, and level $|2\rangle$
is opposite with theirs. Since the two lower levels $|1\rangle$ and
$|2\rangle$ have opposite parity and so
$\langle2|$$\hat{\vec{d}}$$|1\rangle$$\neq0$ where $\hat{\vec{d}}$
is the electric dipole operator. The two upper levels, $|3\rangle$
and $|4\rangle$ have the same parity with
$\langle4|$$\hat{\vec{\mu}}$$|3\rangle$$\neq0$ where
$\hat{\vec{\mu}}$ is the magnetic-dipole operator. As showed in
Figure 1, three electromagnetic fields are introduced to couple the
four states: The electric(\textbf{E})and magnetic(\textbf{B})
components of the probe light(corresponding Rabi frequency
$\Omega_{p}$$=\frac{\vec{E_{P}}\vec{d_{21}}}{\hbar}$ ,
$\Omega_{B}$$=\frac{\vec{B_{P}}\vec{\mu_{43}}}{\hbar}$)interact with
the transitions $|2\rangle$ and$|1\rangle$ as well as$|4\rangle$
and$|3\rangle$, respectively. The incoherent pump field with pumping
rate denoted by $\Gamma$ pumps atoms in level $|1\rangle$ into upper
level $|3\rangle$, and then the atoms decay into metastable level
$|2\rangle$ via rapid nonradiative transitions, whose decay rate is
denoted as $\Gamma_{32}$. The strong coherent field couples states
$|2\rangle$ and $|4\rangle$ with Rabi frequency $\Omega_{c}$. The
equation of the time-evolution for the system is described as
\begin{equation}
\frac{d\rho}{dt}=-\frac{i}{\hbar}[H,\rho]+\Lambda\rho ,
\end{equation}
In which, $\Lambda\rho$ represents the irreversible decay part in
the system. Under the dipole approximation and the rotating wave
approximation the density matrix equations described the system are
written as follows:
\begin{equation}
\dot{\rho_{11}}=\Gamma(\rho_{33}-\rho_{11})+\Gamma_{21}\rho_{22}+\Gamma_{31}\rho_{33}+\Gamma_{41}\rho_{44}+i\Omega_{p}(\rho_{21}-\rho_{12}),
\end{equation}
\begin{equation}
\dot{\rho_{21}}=-(\gamma_{21}+i\Delta_{p})\rho_{21}-i\Omega_{p}(\rho_{22}-\rho_{11})+i\Omega_{c}\rho_{41},
\end{equation}
\begin{equation}
\dot{\rho_{22}}=-\Gamma_{21}\rho_{22}+\Gamma_{32}\rho_{32}+\Gamma_{42}\rho_{44}+i\Omega_{p}(\rho_{12}-\rho_{21})+i\Omega_{c}(\rho_{42}-\rho_{24}),
\end{equation}
\begin{equation}
\dot{\rho_{31}}=-[\gamma_{31}+i(\Delta_{c}+\delta)]\rho_{31}+i\Omega_{p}\rho_{32}+i\Omega_{B}\rho_{41},
\end{equation}
\begin{equation}
\dot{\rho_{32}}=-[\gamma_{32}+i(\Delta_{c}-\Delta_{p}+\delta)]\rho_{32}-i\Omega_{c}\rho_{34}-i\Omega_{p}\rho_{31}+i\Omega_{B}\rho_{42},
\end{equation}
\begin{equation}
\dot{\rho_{33}}=-\Gamma(\rho_{33}-\rho_{11})-\Gamma_{31}\rho_{33}-\Gamma_{32}\rho_{33}+\Gamma_{43}\rho_{44}+i\Omega_{B}(\rho_{43}-\rho_{34}),
\end{equation}
\begin{equation}
\dot{\rho_{41}}=-[\gamma_{41}+i(\Delta_{p}+\Delta_{c})]\rho_{41}+i\Omega_{c}\rho_{21}-i\Omega_{p}\rho_{42}+i\Omega_{B}\rho_{31},
\end{equation}
\begin{equation}
\dot{\rho_{42}}=-(\gamma_{42}+i\Delta_{c})\rho_{42}+i\Omega_{c}(\rho_{22}-\rho_{44})-i\Omega_{p}\rho_{41}+i\Omega_{B}\rho_{32},
\end{equation}
\begin{equation}
\dot{\rho_{43}}=-[\gamma_{43}+i(\Delta_{p}-\delta)]\rho_{43}+i\Omega_{c}\rho_{23}+i\Omega_{B}(\rho_{33}-\rho_{44})
\end{equation}
where the above density matrix elements comply with the conditions:
$\rho_{11}+\rho_{22}+\rho_{33}+\rho_{44}$=1 and
$\rho_{ij}$=$\rho_{ji}^{\ast}$. And $\Gamma_{ij}$(i,j=1,2,3,4)is the
spontaneous emission decay rate from level $|i\rangle$ to level
$|j\rangle$, ignoring the collision broaden effect.
$\gamma_{21}$=$\Gamma_{21}$/2,
$\gamma_{31}$=($\Gamma_{31}+\Gamma_{32}$)/2,
$\gamma_{41}$=($\Gamma_{43}+\Gamma_{42}$)/2,
$\gamma_{42}$=($\Gamma_{43}+\Gamma_{31}+\Gamma_{21}$)/2,
$\gamma_{43}$=($\Gamma_{43}+\Gamma_{42}+\Gamma_{31}+\Gamma_{32}$)/2,
$\gamma_{32}$= ($\Gamma_{32}+\Gamma_{31}+\Gamma_{21}$)/2 are the
decay rates to the corresponding transitions. The detuning of the
fields defined as $\Delta_{p}$=$\omega_{21}$-$\omega_{p}$,
$\Delta_{c}$=$\omega_{42}$-$\omega_{c}$,
$\Delta_{B}$=$\omega_{43}$-$\omega_{P}$, respectively.
$\omega_{ij}$=$\omega_{i}$-$\omega_{j}$ is the transition frequency
of level$|i\rangle$ and $|j\rangle$ (i, j=1,2,3,4) and we have
$\delta$ =$\Delta_{p}$-$\Delta_{B}$. Here we set $\delta$ nonzero in
order to avert the major obstacle mentioned in Ref.\cite{Ref15} in
realizing the predicted effects at a realistic experimental setting.
We solve for the steady-state values of the density matrix elements
$\rho_{43}$ and $\rho_{21}$ in a linear approximation when the probe
field is weak, i.e. $\Omega_{p}$, $\Omega_{B}$$\ll$$\Omega_{c}$,
$\Gamma$, and it can be assumed that almost all the atoms are in the
ground state $|1\rangle$,
\begin{equation}
\rho_{43}=a_{1}\textbf{E}+a_{2}\textbf{B},
\end{equation}
\begin{equation}
\rho_{21}=a_{3}\textbf{E}+a_{4}\textbf{B}
\end{equation}
where the coefficients $a_{1}, a_{2}$ and $a_{3}, a_{4}$ are given
by
\begin{equation}
a_{1}=-\frac{d_{12}Z}{D_{1}D_{2}^{2}D_{3}D_{7}D_{9}[(i\Delta_{p}+\gamma_{21})D_{4}+\Omega_{c}^{2}]\hbar},
\end{equation}
\begin{equation}
a_{2}=\frac{i\Gamma\mu_{34}\Omega_{c}^{2}}{D_{1}D_{2}D_{3}D_{9}\hbar},
\end{equation}
\begin{equation}
a_{3}=\frac{i(D_{4}+\frac{\Gamma\Omega_{c}^{2}}{D_{1}D_{9}})d_{12}}{[\Omega_{c}^{2}+(\gamma_{21}-i\Delta_{p})D_{4}]\hbar},
\end{equation}
\begin{equation}
a_{4}=-\frac{\mu_{34}Z}{D_{1}D_{2}D_{7}D_{9}[(\gamma_{21}-i\Delta_{p})D_{4}+\Omega_{c}^{2}]\hbar}
\end{equation}
and
\begin{equation}
Z=[D_{1}D_{2}\Gamma_{21}+i\Gamma(D_{6}+D_{8}-\gamma_{32}D_{5})]\Omega_{c}^{2},
\end{equation}
\begin{equation}
D_{1}=\gamma_{42}+i\Delta_{c},D_{2}=-i\gamma_{32}+\delta+\Delta_{c}-\Delta_{p},
\end{equation}
\begin{equation}
D_{3}=i\gamma_{43}+\delta-\Delta_{p},
D_{4}=\gamma_{41}+i(\Delta_{c}+\Delta_{p}),D_{9}=\Gamma+\Gamma_{21},
\end{equation}
\begin{equation}
D_{5}=\gamma_{42}+i(\Delta_{c}-\Delta_{p}),
D_{6}=i\gamma_{41}\Delta_{p}+i\gamma_{42}\Delta_{p}-\delta\Delta_{p}-3\Delta_{c}\Delta_{p}+\Omega_{c}^{2},
\end{equation}
\begin{equation}
D_{7}=-i\gamma_{31}+\delta+\Delta_{c},
D_{8}=\gamma_{32}+\gamma_{41}+i(\delta+2\Delta_{c})(\Delta_{c}-i\gamma_{42})(\delta+\Delta_{c})
\end{equation}
The ensemble electric polarization and magnetization of the atomic
medium to the probe field are given by
$\vec{P}=N\vec{d_{12}}\rho_{21}$ and
$\vec{M}=N\vec{\mu_{34}}\rho_{43}$,respectively, where N is the
density of atoms. Then the coherent cross-coupling between electric
and magnetic dipole transitions driven by the electric and magnetic
components of the probe field may lead to
chirality\cite{Ref12},\cite{Ref19}. Substituting equations (13) and
(14) into the formula for the ensemble electric polarization and
magnetization, we have the relations
\begin{equation}
\textbf{P}=\alpha_{EE} \textbf{E}+\alpha_{EB} \textbf{B},\nonumber\\
\textbf{M}=\alpha_{BE} \textbf{E}+\alpha_{BB} \textbf{B}
\end{equation}
where
\begin{equation}
\alpha_{EE}=Nd_{12}a_{3},\nonumber\\\alpha_{EB}=Nd_{12}a_{4},
\end{equation}
\begin{equation}
\alpha_{BE}=N\mu_{34}a_{1},\nonumber\\\alpha_{BB}=N\mu_{34}a_{2}
\end{equation}
Considering both electric and magnetic local field
effects\cite{Ref20}-\cite{Ref21}, \textbf{E }and \textbf{B} in
equation(24)must be replaced by the local fields
\begin{equation}
\textbf{E}_{L}=\textbf{E}+\frac{\textbf{P}}{3\varepsilon_{0}},\nonumber\\\textbf{B}_{L}=\mu_{0}(\textbf{H}+\frac{\textbf{M}}{3})
\end{equation}
As a result, we obtain
\begin{eqnarray}
\textbf{P}=\frac{3\varepsilon_{0}(\mu_{0}\alpha_{BB}\alpha_{EE}-\mu_{0}\alpha_{BE}\alpha_{EB}-3\alpha_{EE})}{\mu_{0}\alpha_{BE}\alpha_{EB}+3\alpha_{EE}-\mu_{0}\alpha_{BB}\alpha_{EE}-9\varepsilon_{0}+3\mu_{0}\varepsilon_{0}\alpha_{BB}}\textbf{E}
\nonumber\\
 +\frac{-9\mu_{0}\varepsilon_{0}\alpha_{EB}}{\mu_{0}\alpha_{BE}\alpha_{EB}+3\alpha_{EE}-\mu_{0}\alpha_{BB}\alpha_{EE}-9\varepsilon_{0}+3\mu_{0}\varepsilon_{0}\alpha_{BB}}\textbf{H}\label{},
\end{eqnarray}
\begin{eqnarray}
\textbf{M}=\frac{9\varepsilon_{0}\alpha_{BE}}{\mu_{0}\alpha_{BB}\alpha_{EE}+9\varepsilon_{0}-\mu_{0}\alpha_{BE}\alpha_{EB}-3\alpha_{EE}-3\varepsilon_{0}\mu_{0}\alpha_{BB}}\textbf{E} \nonumber\\
+\frac{3(\mu_{0}\alpha_{BE}\alpha_{EB}-\mu_{0}\alpha_{BB}\alpha_{EE}+3\varepsilon_{0}\mu_{0}\alpha_{BB})}{\mu_{0}\alpha_{BB}\alpha_{EE}+9\varepsilon_{0}-\mu_{0}\alpha_{BE}\alpha_{EB}-3\alpha_{EE}-3\varepsilon_{0}\mu_{0}\alpha_{BB}}\textbf{H}
\end{eqnarray}
By comparison with equation (1), we obtain the permittivity and the
permeability, and the complex chirality coefficients:
\begin{eqnarray}
\varepsilon=1+\chi_{e}=\frac{6\alpha_{EE}+9\varepsilon_{0}+\mu_{0}[2\alpha_{BE}\alpha_{EB}-\alpha_{BB}(2\alpha_{EE}+3\varepsilon_{0})]}{-3\alpha_{EE}+\mu_{0}[-\alpha_{BE}\alpha_{EB}+\alpha_{BB}(\alpha_{EE}-3\varepsilon_{0})]+9\varepsilon_{0}},
\end{eqnarray}
\begin{eqnarray}
\mu=1+\chi_{m}=\frac{-3\alpha_{EE}+2\mu_{0}[\alpha_{BE}\alpha_{EB}-\alpha_{BB}(\alpha_{EE}-3\varepsilon_{0})]+9\varepsilon_{0}}{-3\alpha_{EE}+\mu_{0}[-\alpha_{BE}\alpha_{EB}+\alpha_{BB}(\alpha_{EE}-3\varepsilon_{0})]+9\varepsilon_{0}},
\end{eqnarray}
\begin{eqnarray}
\xi_{EH}=\frac{9c\mu_{0}\alpha_{EB}\varepsilon_{0}}{-3\alpha_{EE}+\mu_{0}[-\alpha_{BE}\alpha_{EB}+\alpha_{BB}(\alpha_{EE}-3\varepsilon_{0})]+9\varepsilon_{0}},
\end{eqnarray}
\begin{eqnarray}
\xi_{HE}=\frac{9c\mu_{0}\alpha_{BE}\varepsilon_{0}}{-3\alpha_{EE}+\mu_{0}[-\alpha_{BE}\alpha_{EB}+\alpha_{BB}(\alpha_{EE}-3\varepsilon_{0})]+9\varepsilon_{0}}
\end{eqnarray}
In the above, we obtained the expressions for the electric
permittivity and magnetic permeability of the atomic media.
Substituting equations from(30)to(33)into(2), the expression for
refractive index can also be presented. In the section that follows,
we will discuss the negative refractive index of the atomic system
without requiring simultaneously negative both permittivity and
permeability under the appropriate conditions.

\section{Results and discussions}

\begin{figure*}
\centering
\resizebox{0.75\textwidth}{!}{%
  \includegraphics{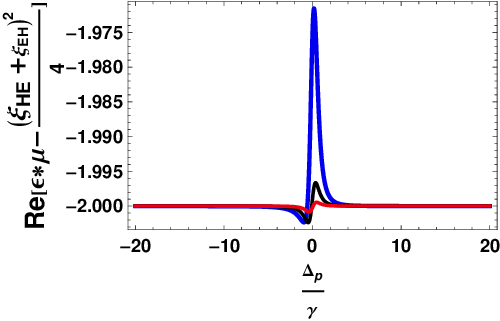}
  \includegraphics{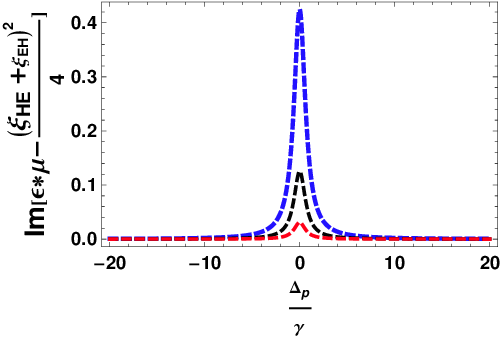}}
\caption{The complex number
[$\varepsilon\mu-\frac{(\xi_{EH}+\xi_{HE})^{2}}{4}$]as a function of
the rescaled detuning parameter $\Delta_{p}/\gamma$ for the three
different sets of the pumping rate of the incoherent pump field and
Rabi frequency of the coherent
field($\Gamma,\Omega_{c}$),respectively.} \label{fig:2}
\end{figure*}

Before doing these calculations, we need to fix several key
parameters such as spontaneous emission rate, wavelength and atomic
density in these equations. In the model configuration, the
transition from$|4\rangle$to$|3\rangle$is magnetic dipole allowed
and others are electrical dipole allowed. The typical value of the
spontaneous emission rate of atomic electric dipole transitions is
the magnitude of $10^{8}$ Hz. The spontaneous emission rate of
atomic magnetic dipole transitions is in general smaller than that
of atomic electric dipole transitions by four magnitude. Thus, in
our numerical calculations, the spontaneous emission rates are
scaled by $\gamma=10^{8}s^{-1}$:
$\Gamma_{43}=\Gamma_{21}\times(\frac{1}{137})^{2}$,
$\Gamma_{21}=1\gamma$, $\Gamma_{41}=0.1\gamma$,
$\Gamma_{42}=0.9\gamma$ ,$\Gamma_{31}=0.3\gamma$,
$\Gamma_{32}=0.2\gamma$. The typical optical wavelength for the
transitions $|4\rangle$$\rightarrow$$|3\rangle$ and
$|2\rangle$$\rightarrow$$|1\rangle$ is selected to be 600 nm[22,23].
The dipole moments $d_{12}$ and $\mu_{34}$ are estimated by the
relation$\sqrt{3\hbar\Gamma_{ij}\lambda^{3}/8\pi^{2}}$. In the
present calculations, we choose the density of atoms N to be
$5\times10^{22}m^{-3}$. The level configuration shown in figure 1
may be realized in the trivalent positive ions $E^{3+}_{r}$ doped
into calcium fluorophosphate at room temperature,which has abundant
energy levels, various electric magnetic transitions and high
density\cite{Ref24},\cite{Ref25}. The four levels
$|1\rangle$,$|2\rangle$,$|3\rangle$, and $|4\rangle$in Figure 1 may
correspond to the energy level configuration
$I_{15/2}^{4}$,$I_{13/2}^{4}$, $I_{9/2}^{4}$ and $I_{11/2}^{4}$ of
the rare earth ion $E^{3+}_{r}$ doped in calcium fluorophosphate,
respectively.The detuning of the strong coherent field is set as
$-5\times10^{-3}$$\gamma$, and the difference of electric and
magnetic coupling $\delta=-10^{-3}$$\gamma$. In Ref.\cite{Ref15},
the probe field hypothesized resonance with the electric and
magnetic transitions is considered a major obstacle in realizing the
predicted effects at a realistic experimental setting, as it is not
straightforward to find a system with two states fitting the
condition. And in our scheme the obstacle dissolve because the
electric and magnetic transitions are set to be off-resonant.

The pumping rate of the incoherent pump field $\Gamma$ and the Rabi
frequency of the coherent field $\Omega_{c}$ are set as parameter
group ($\Gamma,\Omega_{c}$) in the following discussion. And the
blue, black and red curves in all the figures are corresponding to
the three different parameter groups,(5$\gamma$, 20$\gamma$),
(50$\gamma$, 10$\gamma$) and (25$\gamma$, 5$\gamma$), respectively.
In figure 2, the complex number
[$\varepsilon\mu-\frac{(\xi_{EH}+\xi_{HE})^{2}}{4}$] is shown
against the rescaled detuning parameter $\Delta_{p}$/$\gamma$. It is
observed that the sign of the real parts of the complex number is
minus with the different parameter groups, while its imaginary part
remains positive sign when the parameter groups vary with the three
different values. At the resonant point the peak values are
decreasing with the diminish of the Rabi frequency $\Omega_{c}$.
With regard to a complex number, its argument is in the second
quadrant of cartesian coordinates system when it has minus real part
and positive imaginary part. And the principal argument value of the
square root
[$\sqrt{\varepsilon\mu-\frac{(\xi_{EH}+\xi_{HE})^{2}}{4}}$] is a
half of the complex number
[$\varepsilon\mu-\frac{(\xi_{EH}+\xi_{HE})^{2}}{4}$]. Because the
argument of
[$\sqrt{\varepsilon\mu-\frac{(\xi_{EH}+\xi_{HE})^{2}}{4}}$] is
constituted by the principal argument angle and $k\pi$, its phase is
in the third quadrant of cartesian coordinates when k is assigned to
an odd number. Thus the real part of the first part of the
refractive index gets negative value, because its first part is
[$\sqrt{\varepsilon\mu-\frac{(\xi_{EH}+\xi_{HE})^{2}}{4}}$]. In
equation(2), the real part of the refractive index gets negative
value if its first part has sufficiently large magnitude.

\begin{figure*}
\centering
\resizebox{0.75\textwidth}{!}{%
  \includegraphics{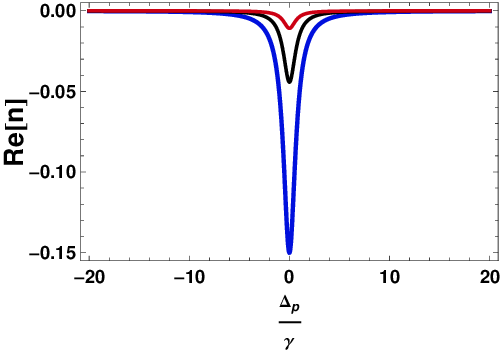}
  \includegraphics{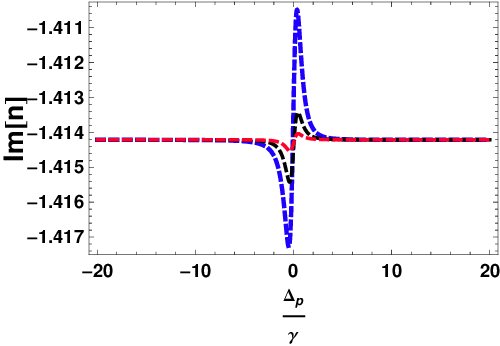}}
\caption{The real and imaginary parts of the refractive index as a
function of the rescaled detuning parameter $\Delta_{p}/\gamma$ for
the three different sets of pumping rate of the incoherent pump
field and Rabi frequency of the coherent
field($\Gamma,\Omega_{c}$),respectively.}
\label{fig:3}
\end{figure*}

\begin{figure*}
\centering
\resizebox{0.75\textwidth}{!}{%
  \includegraphics{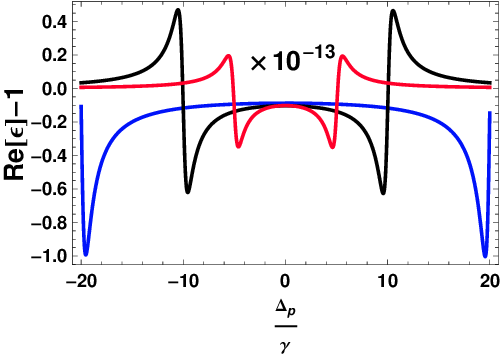}
  \includegraphics{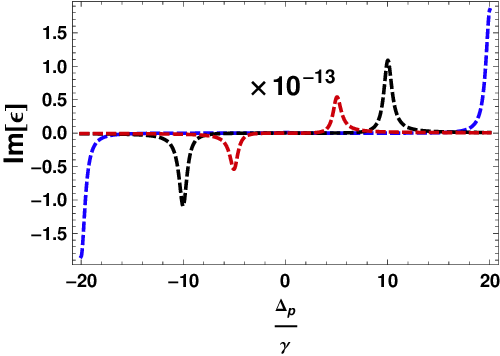}
  }
\caption{The real and imaginary parts of the permittivity as a
function of the rescaled detuning parameter $\Delta_{p}/\gamma$ for
the three different sets of pumping rate of the incoherent pump
field and Rabi frequency of the coherent
field($\Gamma,\Omega_{c}$),respectively.}
\label{fig:4}
\end{figure*}

In figure 3, the refraction index is plotted for the different three
parameter groups. It's observed that the real and imaginary parts of
the refractive index are negative and their the largest absolute
value present to the resonant point when $\Omega_{c}$=20$\gamma$. In
figure 4, the permittivity is shown against the rescaled detuning
parameter $\Delta_{p}$/$\gamma$. As shown in figure 4, the rescaled
real parts of permittivity ($Re[\varepsilon]-1$) have the same order
of magnitude $10^{-13}$, and the same value -0.1 in the near of
magnitude ($10^{-13}$) is far less than the number 1. We note that
the imaginary part of permittivity has the same order of magnitude
with its real part. Beside the absorption peaks and the gain vales
at the symmetry points, the transparent windows appear in the curves
of the imaginary part of permittivity. Which concludes that the
maximum intensity of the coherent field produces the widest
transparent window. From the figure 5, we observed the rescaled real
part permeability ($Re[\mu]-1$) has a more less order of magnitude
($10^{-24}$) than that of the rescaled real parts of permittivity
($Re[\varepsilon]-1$). And the imaginary part $Im[\mu]$ gets the
magnitude of ($10^{-25}$). As shown in figure 5, the response from
the external fields only concentrates in the near resonant region.
And the strongest coherent field corresponds the largest
permeability. The magnitude of the rescaled real parts of
permittivity ($Re[\mu]-1$) is $ Re[\mu]\approx 1 $. Thus the
electric and magnetic susceptibilities are weak because of
$Re[\varepsilon]\approx 1$ and $Re[\mu]\approx 1$. So we obtain the
negative refraction completely resulted from the electromagnetically
induced chirality in the scheme.

\begin{figure*}
\centering
\resizebox{0.75\textwidth}{!}{%
  \includegraphics{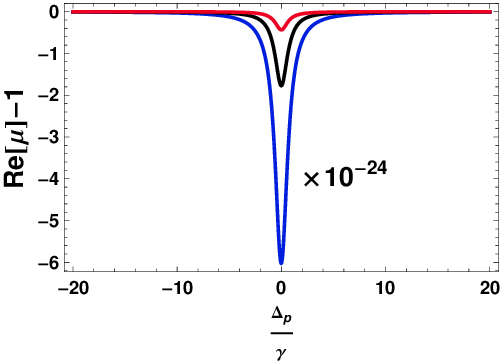}
  \includegraphics{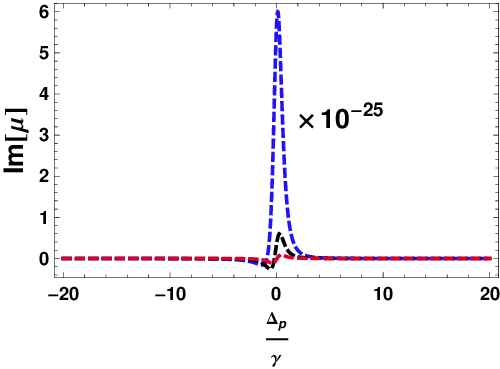}}
\caption{The real and imaginary parts of the  permeability as a
function of the rescaled detuning parameter $\Delta_{p}/\gamma$ for
the three different sets of pumping rate of the incoherent pump
field and Rabi frequency of the coherent
field($\Gamma,\Omega_{c}$),respectively.}
\label{fig:5}
\end{figure*}

\section{Conclusion}

In summary, electromagnetic chirality-induced negative refraction in
a four-level atomic medium is obtained by the algebraic analysis to
the argument of the refractive index for one circular polarization.
When two unequal transition frequencies responding to the electric
and magnetic transitions of the probe field and the parameter
groups($\Gamma,\Omega_{c}$) getting different values, the atomic
system shows negative refraction without requiring both electric
permittivity and magnetic permeability to be simultaneously
negative. Compared with the method of realizing negative
fraction\cite{Ref18},\cite{Ref22}, our algebraic analysis scheme
seems to reduce the difficulty and provide more possibilities. It
may possibly give a novel approach to obtain the desired material
with negative refractive index by electromagnetic
chirality-inducing.

\end{document}